\documentclass[aps,prl,floats,twocolumn,epsf]{revtex4-1}
\usepackage{graphicx}
\usepackage{epsfig}

\def \beq{\begin{equation}}
\def \eeq{\end{equation}}
\def \beqarr{\begin{eqnarray}}
\def \eeqarr{\end{eqnarray}}
\def \bspt{\begin{split}}
\def \espt{\end{split}}
\def \bef{\begin{figure}}
\def \enf{\end{figure}}

\begin{document}


\title{Adiabatic Cooling with Non-Abelian Anyons}

\author{G. Gervais$^{1}$ and Kun Yang$^{2}$}

\affiliation{$^{1}$Department of Physics, McGill University,
Montreal, H3A 2T8, CANADA}

\affiliation{$^{2}$National High Magnetic Field Laboratory
and Department of Physics, Florida State University, Tallahassee,
Florida 32306, USA}

\date{\today}

\begin{abstract}

We show in this Letter that the ground state degeneracy associated with the presence of non-Abelian anyons can be probed by using an adiabatic cooling process based on the non-Abelian entropy. In particular, we show that when the number of such anyons is increased adiabatically at sufficiently low temperatures, the non-Abelian liquid undergoes cooling, whereas heating occurs in the Abelian case. Estimates are provided for the cooling power produced by the non-Abelian anyon refrigerator, and its implementation in non-Abelian fractional quantum Hall liquids is discussed.

\end{abstract}


\maketitle

{\it Introduction:} Recently there has been very strong interest in unusual fractional quantum Hall (FQH) states whose quasiparticle excitations obey non-Abelian statistics\cite{nayak08}. Such interest is partially driven by the potential of using such non-Abelian quasiparticles, or non-Abelian anyons, for quantum information storage and processing in an
intrinsically fault-tolerant
fashion~\cite{nayak08,kitaev03,dassarma05}. At this time the most promising candidate for non-Abelian statistics is the FQH state at filling factor $\nu_0=5/2$\cite{willett87,pan,Miller,Dean}, in which the electrons in the half-filled first excited Landau level may condense into the Moore-Read (MR, or Pfaffian) state\cite{moore91} or its particle-hole conjugate (anti-Pfaffian state)\cite{lee07,levin07}.  Theoretical support for the Pffafian or anti-Pfaffian state as an explanation for the FQH state at $\nu_0=5/2$ has come from a variety of numerical calculations\cite{morf,rh,wan06,wan08,feiguin,peterson,moller,wang,storni}.\\

Experimentally, the quasiparticle charge $e^*$ has been measured via tunneling between opposite edges across a constriction\cite{dolev,radu}, as well as interference pattern between two separate constrictions\cite{willett08}, and found to be consistent with theoretical predictions. Particularly exciting is the recent observation of interference patterns which alternate between those of quasiparticles with $e^*=e/4$ and $e^*=e/2$\cite{willett09}; this can be naturally understood in terms of the ``even-odd effect" associated with the non-Abelian nature of quasiparticles with $e^*=e/4$\cite{stern06,bonderson06}, combined with the Abelian nature of quasiparticles with $e^*=e/2$\cite{wan08}. While the on-going edge interference experiments are encouraging, complications associated with the edge, such as edge reconstruction and coupling between bulk quasiparticles and the edge, may make their interpretation difficult. 
It would thus be highly desirable to have alternative methods that can directly probe non-Abelian anyons in the {\em bulk}. To this end
recent proposals have been put forward to probe non-Abelian quasiparticles using bulk thermoelectric measurements, including thermopower\cite{yh09} and temperature dependence of electrochemical potential and magnetization\cite{cooperstern}. Such bulk probes are complimentary to the edge probes as instead of measuring the braiding matrix elements, they measure the entropy associated with ground state degeneracy due the presence of non-Abelian quasiparticles, thus allowing for the determination of their quantum dimensions\cite{yh09}. Very recently thermopower has been measured near $\nu=5/2$\cite{chikering}. While for technical reasons the temperature ($T\gtrsim 82 mK$) in that work may not have reached the low-$T$ regime necessary to test theoretical predictions\cite{yh09} quantitatively, the results do seem to indicate that non-Abelian entropy, if present, would represent a sizeable fraction of the total entropy of the system. This certainly gives us hope that the non-Abelian entropy may indeed be accessible in bulk  measurements.\\


{\it Proposal:} Here,  we propose another bulk probe of the non-Abelian entropy that is very simple and can be summarized as follows: The total entropy of the 2D electronic system is the sum of the (temperature-independent) non-Abelian entropy $S_D\propto N_q$, where $N_q$ is the number of non-Abelian quasiparticles, and other (normal) sources of entropy:
\beq
S_{tot}=S_D+ S_n(T).
\label{stot}
\eeq
Here $S_n(T)$ is the entropy due to normal excitations of the system, which is a monotonically increasing function of temperature $T$. When one changes $N_q$ {\em adiabatically} by  either changing the magnetic field $B$ or the electron density $n_e$, the total entropy $S_{tot}$ does not change; as a result any change in $S_D$ must be compensated for by an equal and opposite change in the normal entropy $S_n(T)$. We therefore expect an adiabatic increase in $N_q$ will {\em decrease T}  at sufficiently low temperature, {\it i.e.}  when $S_D$ dominates over $S_n$. It should therefore  be possible {\it in principle} to control the temperature by controlling $N_q$ through the manipulation of either the electron density $n_e$ or the magnetic field, $B$.  The rest of this paper is devoted to a detailed analysis of this simple effect.\\

{\it Analysis:} A key property of non-Abelian statistics is the appearance of a ground state degeneracy $D$ that grows (up to an $O(1)$ prefactor) exponentially with the number of quasiparticles present in the system, $N_q$, when their positions are fixed:
\beq
D\sim d^{N_q},
\label{degeneracy}
\eeq
where $d > 1$ is the quantum dimension\cite{nayak08} of the quasiparticle. For the non-Abelian quasiparticles in the MR Pfaffian or anti-Pfaffian state, $d=\sqrt{2}$. We will use them as the primary examples of our discussion below, although essentially all of our discussions apply to other non-Abelian FQH states. Such degeneracy results in a ground state entropy
\beq
S_D=k_B\log D = k_B N_q\log d +O(1),
\label{entropy}
\eeq
where $k_B$ is the Boltzmann constant; {\em i.e.}, each quasiparticle carries entropy $k_B\log d$. In principle, there exists very weak coupling among the quasiparticles that can lift the ground state degeneracy\cite{anyonchain,baraban,bonderson09}; however such coupling vanishes exponentially as a function of the distance between quasiparticles. Thus the non-Abelian entropy formula Eq. (\ref{entropy}) remains valid for as long as
\beq
T \gg T_0,
\label{T0}
\eeq
where $T_0$ is the temperature scale associated with quasiparticle couplings and is estimated to be inaccessibly low\cite{yh09}; we can therefore assume Eq. (\ref{T0}) to be always satisfied.

In a uniform system, the number of quasiparticles at low temperatures is proportional to the deviation of the magnetic field $B$ from the value $B_0$ at the center of the FQH plateau, where the filling fraction is equal to the ideal value $\nu_0$:
\beq
N_q=|(e/e^*)(B-B_0)/B_0|N_e= (e/e^*)(\Delta B/B_0)N_e,
\label{quasiparticlenumber}
\eeq
where $N_e$ is the number of electrons in the system, and $\Delta B= |B-B_0|$. As a result the non-Abelian entropy,
\beq
S_D= (e/e^*)(\Delta B/B_0)N_e k_B\log d
\label{SD}
\eeq
grows linearly with $\Delta B$. We note that $\Delta B$ can be varied by changing either $B$ or the electron density $n_e$, which in turn changes $B_0$.\\

We now turn our discussion to the normal component of the entropy $S_n(T)$, which for $T$ sufficiently lower than the quasiparticle gap is dominated by the positional degrees of freedom of the existing quasiparticles. At $T$ sufficiently low, and  at low density of quasiparticles, we expect
the quasiparticles  to form a Wigner crystal, with positional entropy coming from thermally excited magnetophonons. Treating the quasiparticles as point particles with charge $e^*$ moving in the magnetic field $B$, which is justified in the dilute limit, they form a triangular lattice with lattice spacing
\beq
a=l_B\left[{4\pi\over \sqrt{3}\nu_0}{e^*\over e}{B_0\over \Delta B}\right]^{1/2},
\eeq
where $l_B\equiv \sqrt{\frac{\hbar c}{eB}  }$ is the electron magnetic length. Using the known magnetophonon spectrum of that system\cite{fukuyama}, we obtain the quasiparticle Wigner crystal magnetophonon spectrum at long wave length or small wavevector $k$:
\beq
E_{phonon}(k)=E_0(\Delta B)(ka)^{3/2},
\eeq
where the energy scale $E_0(\Delta B)$ is
\beq
E_0(\Delta B) \approx 0.071{e^2\over\epsilon l_B}\sqrt{{e\over e^*}}\left[{\nu_0\Delta B\over B_0}\right]^{3\over 2}.
\label{E0}
\eeq
At sufficiently low temperatures where $T \lesssim E_0/k_B$, we obtain for the normal entropy
\beq
S_n(T)\approx \alpha N_e k_B{e\over e^*}{\Delta B\over B_0}\left[{k_B T\over E_0(\Delta B)}\right]^{4\over 3},
\label{sn}
\eeq
where $\alpha \approx 0.27$.
Considering now an  {\em adiabatic} change of $\Delta B$ producing a net zero change in the total entropy,
\beq
{d S_{tot}\over d\Delta B}={d S_D\over d\Delta B}+{d S_n\over d\Delta B}=0;
\eeq
 we obtain as a result a net temperature change
\beq
{dT\over d\Delta B}={3T\over 4\Delta B}\left\{1-{\log d\over \alpha}\left[{E_0(\Delta B)\over k_B T}\right]^{4\over 3}\right\};
\eeq
and importantly  we find a {\it net negative} temperature change, or cooling effect
\beq
{dT\over d\Delta B} < 0
\eeq
for as long as $T < T^*$, where
\beq
T^*={E_0(\Delta B)\over k_B}\left(\log d\over \alpha\right)^{3\over 4}.
\eeq
Furthermore, in the limit $T\rightarrow 0$,
\beq
{dT\over d\Delta B} \propto  -T^{-1/3}
\eeq
the cooling effect {\em diverges}.  It is illuminating to compare these results with the case of {\em Abelian} quasiparticles  with quantum dimension $d=1$ and $S_D=0$; in this case we obtain instead
\beq
{dT\over d\Delta B}= {3T\over 4\Delta B} > 0.
\eeq
Changing the magnetic field $\Delta B$ adiabatically has the {\em opposite} effects on $T$ for Abelian and non-Abelian quasiparticles in the same low temperature regime,  $T < T^*$. These are the central results of this work.\\

{\it Discussion:} We now discuss in more details the stringency of the conditions that need to be met in order to 
observe the non-Abelian cooling effect. An important temperature scale to be considered is the melting temperature $T_m$ of the Wigner crystal. Its classical value is  given by a small fraction of the Coulomb interaction energy between quasiparticles,
\beq
k_BT_m = {1\over \Gamma}{(e^*)^2\over\epsilon l_B}\left[{\nu_0\Delta B\over 2B_0}{e\over e^*}\right]^{1\over 2},
\label{freezing}
\eeq
where $\Gamma\approx 137$\cite{grimes,gann,morf79}. Strictly speaking, our quantitative analysis above holds for  the crystalline  phase when $T < T_m$. However, we expect our results to hold qualitatively or even semi-quantitatively for $T_m < T < T^*$ (when $T_m < T^*$), if melting is continuous or very weakly first order. This is because in this case the liquid state that results from melting is expected to have strong short-range crystal order, and its positional entropy remains close to Eq. (\ref{sn}) as along as $T \lesssim E_0/k_B$. Experiment\cite{grimes} as well as numerical simulation of classical Coulomb system suggest that the melting transition is indeed continuous\cite{gann,morf79}.\\

In addition to melting, the ubiquitous presence of disorder will also give corrections to Eq. (\ref{sn}). In particular, a quasiparticle Wigner crystal is expected to be pinned by disorder, as is experimentally observed in the microwave transport of integer quantum Hall states\cite{Lloyd}. This pinning is in essence  what gives rise to the FQH plateau in the first place and in general increases the magneto-phonon energy, and thus is likely to  {\em suppress} the entropy $S_n(T)$ [Eq. (\ref{sn})]. This will most likely {\em increase} $T^*$.\\

{\it Experimental Prospect:}  Focusing on GaAs-based two dimensional electron gases,  the cooling power of the non-Abelian adiabatic refrigerator  can be estimated from Eq. (\ref{sn}), using  the thermodynamic relation

\beq
\dot{Q}=C(T)\frac{dT}{dt}=C(T)\frac{dT}{d\Delta B}\frac{d\Delta B}{dt},
\eeq
where the specific heat of the 2D electron gas is given by

\beq
C(T)= T\frac{dS_{n}}{dT}=\frac{4}{3} \alpha N_{e}k_{B}\frac{e}{e^{*}}\frac{\Delta B}{B_{0}} \left[ \frac{k_{B}T}{E_{0}(\Delta B)}\right]^{\frac{4}{3}}.
\eeq

Using the sample of Ref. \onlinecite{chikering} (as an example) with  the 5/2 FQH state located at a magnetic field  $B_0\simeq 5 T$ and with a number of electrons $N_{e}\simeq 3\times 10^{11}$, the magnetic length at this field is $l_B\simeq 10$ $nm$, and at the edge of the plateau $\Delta B/B_0\simeq 1/100$, indicating the quasiparticles form a (pinned) Wigner crystal up to that point, at low temperatures. Using the dielectric constant $\epsilon=13$ and $e^*/e=1/4$, we obtain a melting temperature $T_m\simeq 11$ $mK$ and $T^*\simeq 75$ $mK$ at the plateau edge ($\Delta B=0.050$ $T$). These set the (rough) temperature scales below which the predictions presented above are expected to hold. Below these temperatures, the cooling power can be estimated  when the quasiparticle density is changed by a magnetic field, and swept (for example) at a rate $d\Delta B/dt=0.010$ $T/s$, or equivalently through a $\sim0.2 \%/s$ change in electron density via gating. With these values, we estimate  the cooling power at $T=10$ $mK$ for the non-Abelian case to be $\dot{Q}_{n-Abelian}\simeq -0.1$ $fW$ at $\Delta B\simeq 0.05$ $T$, and a heat production at a rate  $\dot{Q}_{Abelian}\simeq 0.01 $ $fW$ for Abelian quasiparticles. Albeit small, these power values compare well with, and in the non-abelian liquid remains above the joule heating produced by a typical 1 $nA$ measurement current flowing in an electron gas with  $R_{xx}\simeq 10$ $\Omega$ and producing $\dot{Q}_{joule}\simeq 0.01$ $fW$. For cooling to be observable, the cooling power of the non-Abelian refrigerator should be of order, or larger than other sources of heat leakage coming from the environment, {\it i.e.} $\dot{Q}_{n-Abelian}\gtrsim  \dot{Q}_{leak}$. These sources of leakage  may come from the lattice, radiation and from the electrical contacts used to flow electrical currents in the electron gas. \\

It is well-known that cooling an electron gas down to the lowest temperatures is challenging experimentally due to the freezing of  phonons in GaAs below $\sim$50 $mK$. Albeit difficult, electron cooling via the electrical leads has nevertheless been achieved down to $\sim4$ $mK$\cite{pan}. This thermal disconnect effectively isolates the electron gas from the lattice, a highly desirable situation for our proposal.  The thermal relaxation time of a two dimensional electron gas has been studied recently, and observed to grow with decreasing $T$ and increasing $B$, and observed to be as long as minutes at $T\sim 80$ $mK$ and $B\sim 5T$ in high quality samples\cite{chikering}. While this can be a problem in  thermal experiments, it becomes an asset in the experiment proposed here since the adiabatic condition can be reached rather easily at low $T$. In implementing this proposal, one can also vary $\Delta B$ adiabatically by either changing $B$ while fixing $n_e$, or by changing $n_e$ (and thus changing $B_0$) while holding $B$ fixed. In samples where $n_e$ can be varied by gating, varying  $n_e$ may be
better suited since it is less prone to the eddy currents induced by a time-varying magnetic field which could dissipate heat in the electron gas. \\

The adiabaticity condition  also requires that entropy is not flowing in and out of the electron gas while the
quasiparticle density is varied. This can be achieved in principle by thermally disconnecting  the electron gas from the rest of the cryo-environment  with a heat switch made out of a superconducting material that  does not carry entropy at low temperatures. Equally important  is the experimental determination of the electronic temperature, a difficult task that can be achieved possibly {\it in situ} by measuring the longitudinal resistance  $R_{xx}$ at the edge of the plateau. \\

{\it Conclusion:} We have proposed a scheme for, and have studied theoretically the adiabatic cooling of non-Abelian quasiparticles in a bulk fractional Hall quantum liquid. Importantly, our analysis shows that adiabatically  increasing the number of quasiparticles produces a cooling effect for non-Abelian states, whereas the same process leads to heating in the Abelian case. Estimates for the cooling power, and of various temperature scales, suggest that adiabatic cooling in non-Abelian quantum Hall liquids might be within experimental reach. If confirmed by experiments, this would provide a verification of  the ground state degeneracies predicted to exist in the presence of non-Abelian quasiparticles.\\

While we have focused on FQH states, the ideas discussed here also apply to other systems that support non-Abelian anyons. For example, vortices of the $p_x+ip_y$ superconductors or superfluids are expected to carry with them Majorana fermions (or Ising anyons) which give rise to ground state degeneracy and entropy in a manner identical to the 5/2 FQH state\cite{nayak08}; we would therefore expect a similar cooling effect 
to occur (at sufficiently low temperatures) when an adiabatic increase in magnetic field is generated. \\

{\it Acknowledgments:}  This work has been supported by the Natural Sciences and Engineering Research Council of Canada (NSERC), the Canadian Institute for Advanced Research (CIFAR), FQRNT (Qu\'{e}bec), and NSF grant DMR-0704133. We  would like to thank B.I. Halperin for illuminating discussions, and
B.A. Schmidt for assistance during the preparation of this manuscript.

\end{document}